\documentstyle[12pt]{article}
\begin{document}
\rightline{IFT-P.022/92}
\rightline{IFUSP/P-996}
\rightline{July 1992}
\hfill{hep-ph/9207262}
\vskip 1.5cm
{\large \bf \center Comment on the $\tau$ decay puzzle\\}
\vskip 2cm
{\center C.O. Escobar$^{a}$, O.L.G. Peres$^{b}$, V.
Pleitez$^{b}$  and
R. Zukanovich Funchal$^{a}$\\
\vskip .8cm
$^{a}$ Instituto de F\'\i sica da Universidade de S\~ao Paulo,\\
01488-970--S\~ao Paulo, SP, Brazil.\\
\vskip .5cm
$^{b}$ Instituto de F\'\i sica Te\'orica \\
Universidade Estadual Paulista, Rua Pamplona, 145 \\
01405-900--S\~ao Paulo, SP, Brazil.\\}
\vskip 1cm
{\center Abstract\\}
\vskip .5cm
We analyze the current data on $\tau$-lepton decays and show that they
are consistent with the Standard Model.

\newpage

There has been recently great interest on possible problem with
decays of the $\tau$-lepton. These range from deviations of universality
{}~\cite{xm} to problems with branching ratios into particular decay
channels~\cite{ss,rs}.

In this note we point out that, provided an adequate analysis is
adopted when comparing theoretical predictions with experimental
data, there is no such problem.

First of all it should be stressed that the only theoretical
prediction that can be actually made is the prediction for the
partial rate into a particular channel $(\Gamma^\tau_i)$, where
$\Gamma^\tau_i$ is the $\tau$ partial width of the decay into the $i$
charged particle. It is not possible to theoretically predict
branching ratios if we do not know all the decay channels. This is
precisely the case of the $\tau$ lepton.

Now that we have defined what is theoretically predictable, it is
rather clear what should be the procedure for comparison with
experimental data. We should compare theory $(\Gamma^\tau_i)$ with
those experiments which measure the same quantity, i.e., experiments
having both measurements for branching ratios $(B.R)$ and
$\tau$ lifetime $(\tau_\tau)$. We will use the notation
$B^\tau_i=B.R.(\tau\rightarrow
i+X)$ where $i$ means the charged particle (lepton or hadron) and $X$
denotes the corresponding neutrinos. We do not adopt the
procedure of averaging the existent world data since this
ignores the fact that not all experiments have
measured both quantities. It is more sound to make comparison on a
experiment by experiment basis, which is what we now present.

Let us start by the theoretical result for the partial width from the
paper of Marciano~\cite{m,ms}, which includes radiative corrections,
using the current data from Ref.~\cite{pdg}
\begin{eqnarray}
\Gamma(\tau\rightarrow
e^-\nu_\tau\bar\nu_e)&=&(4.11^{+0.03}_{-0.04})\times10^{-13}\mbox{GeV},
\nonumber \\
\Gamma(\tau\rightarrow \mu^-\nu_\tau\bar\nu_\mu)&=&
(4.00^{+0.03}_{-0.04})\times 10^{-13}\mbox{GeV},
\nonumber \\
\Gamma(\tau\rightarrow
\pi\nu_\tau)&=&(2.55\pm0.07)\times10^{-13}\mbox{GeV} ,\nonumber \\
\Gamma(\tau\rightarrow
K\nu_\tau)&=&(1.66\pm0.04)\times10^{-14}\mbox{GeV}, \nonumber \\
\Gamma(\tau\rightarrow h\nu_\tau)&=&(2.71\pm0.07)\times10^{-13}\mbox{GeV}.
\label{1}
\end{eqnarray}
The asymmetric errors come from the errors in the $\tau$ mass $(m_\tau)$ in
Ref.~\cite{pdg},
$m_\tau=1784.1^{+2.7}_{-3.6}\mbox{MeV}$. These
results are compatible with calculations using the preliminary more
precise measurement from BES experiment at BEPC in Beijing,
$m_\tau=1776.9\pm0.4\pm0.3$ MeV~\cite{bb}.

With the above theoretical values the following ratios can be
computed:
\begin{eqnarray}
\Gamma^\tau_{\mu e}&=&0.973^{+0.010}_{-0.014}\, ,\nonumber \\
\Gamma^\tau_{he}&=& 0.660^{+0.017}_{-0.018}\,, \nonumber \\
\Gamma^\tau_{h\mu}&=& 0.678^{+0.017}_{-0.018}\,,
\label{2}
\end{eqnarray}
where we have defined $\Gamma^\tau_{ij}=\Gamma^\tau_i/\Gamma^\tau_j$
and $\Gamma^\tau_h=\Gamma^\tau_\pi+\Gamma^\tau_K$.

If instead we want to use all available data we compare the ratios of
branching ratios~(\ref{2}), since this is also a clear theoretical
prediction, with the ratios of experimental ratios
$R_{ij}=B^\tau_i/B^\tau_j$. This comparison is displayed in
Fig. 1. We see that there is also no conflict between two
standard deviations.

The experimental results~\cite{a,ar} are presented in
Table~\ref{t1}.

We compare the theoretical predictions (\ref{1}) with the data
$B^\tau_i/\tau_\tau$ in Table~\ref{t2}. For those experiments having
measurements of the branching ratios and lifetime we clearly see from
Table~\ref{t2} and Eqs.~(\ref{1}) that there is no conflict
between theory and experiments within two standard deviations.

The ratio between the leptonic and the hadronic width is
particularly useful as a theoretical prediction since this ratio
does not depend on the $\tau$ lifetime but is dependent on the $\tau$
mass as $m_\tau^2$. The agreement we just pointed out makes less
important the issue of $\tau$ mass determination as
stressed in Ref.~\cite{m}

Let us finally remark that mixing with a fourth generation although
solving the $\tau$ lifetime problem~\cite{ss,rs}, will not
solve discrepancies in individual branching ratios~\cite{m}.

We conclude that the present experimental data on the $\tau$ lepton
decays are compatible with the Standard Model.

\vskip 2cm
\noindent
{\bf Acknowledgements}

We are very gratefull to Funda\c c\~ao de Amparo \`a Pesquisa do
Estado de S\~ao Paulo (FAPESP) (R.Z.F.), Coordenadoria de Aperfei\c
coamento de Pessoal de N\'\i vel Superior (CAPES) (O.L.G.P.) for full
financial support and Con\-se\-lho Na\-cio\-nal de
De\-sen\-vol\-vi\-men\-to Cien\-t\'\i \-fi\-co e
Tec\-no\-l\'o\-gi\-co (CNPq) (V.P.) for partial financial support.

\newpage

\centerline{\bf FIGURE CAPTIONS}
\vskip 0.5 true cm
\noindent
{\bf Fig. 1}  Comparison of the theoretical (thick line) and
experimental results for: $(a)$ ratio of the branching ratios for
hadronic and leptonic decays, $R_{h\mu}=B^\tau_h/B^\tau_\mu$
(continuous line) and $R_{he}=B^\tau_h/B^\tau_e$ (dashed line); $(b)$
ratio of the leptonic branching ratios, $R_{\mu
e}=B^\tau_\mu/B^\tau_e$ (dotted line).

\newpage

\begin{table}
\begin{tabular}{|c|c|c|c|c|} \hline
Experiment & $\tau_\tau\,(10^{-13} sec)~\cite{a}$& $B^\tau_e\,
(\%)~\cite{ar}$  &
$B^\tau_\mu\,(\%)~\cite{ar}$  & $B^\tau_h  \,(\%)~\cite{ar}$\\ \hline
$ALEPH$    & $2.95\pm0.10\pm0.05$   &18.09$\pm0.45\pm0.45$
&17.35$\pm0.41\pm0.37$ &12.55$\pm0.55$
\\ \hline
$OPAL$     &3.08$\pm0.13$  &17.4$\pm0.5\pm0.4$    &16.8$\pm0.5\pm0.4$
&12.1$\pm0.7\pm0.5$
\\ \hline
$L3$       &3.09$\pm0.23\pm0.30$ &17.7$\pm0.7\pm0.6$
&17.5$\pm0.8\pm0.5$
&---------------
\\ \hline
$DELPHI$   &3.14$\pm0.23\pm0.04$
& $18.6\pm0.8\pm0.6$ &$17.4\pm0.7\pm0.6 $  & $11.9\pm0.7\pm0.7$
\\ \hline
$ARGUS$    &2.95$\pm0.14\pm0.11$ &17.3$\pm0.4\pm0.5$
&17.2$\pm0.4\pm0.5$
&11.7$\pm0.6\pm0.8$
\\ \hline
$CLEO$     &3.25$\pm0.14\pm0.18$  &19.2$\pm0.4\pm0.6$
&----------------&---------------
 \\ \hline
$CELLO$    &--------------&18.4$\pm0.8\pm0.4$  &17.7$\pm0.8\pm0.4$
&12.3$\pm0.9\pm0.5$ \\  \hline
\end{tabular}
\caption{ Experimental data for $\tau$-decays.}
\label{t1}
\end{table}


\begin{table}
\begin{tabular}{|c|c|c|c|} \hline
Experiment &$
R^\tau_e/\tau_\tau\times10^{-13}\mbox{GeV}$~\cite{a,ar}&
$R^\tau_{\mu}/\tau_\tau\times10^{-13}\mbox{GeV}~\cite{a,ar}$
&$R^\tau_{h}/\tau_\tau\times10^{-13}\mbox{GeV}~\cite{a,ar}$
\\ \hline
$ALEPH$   & $4.04\pm0.21$ &$ 3.87\pm0.19$ & $2.80\pm0.16$
\\ \hline
$OPAL$    &3.72$\pm0.21$            &3.59$\pm0.20$
&2.59$\pm0.21$
\\ \hline
$L3$      &3.77$\pm0.50$            &3.73$\pm0.50$
&---------------------
      \\ \hline
$DELPHI$  &$ 3.90\pm0.37$ & $3.65\pm0.35$
&$2.50\pm0.29$
\\ \hline
$ARGUS $  &$3.86\pm0.27$            &3.84$\pm0.28$
&2.61$\pm0.27$
\\ \hline
$CLEO$    &3.89$\pm0.31$            &------------------------
&---------------------
\\ \hline
\end{tabular}
\caption{Experimental partial widths, with the same notation of Table 1.}
\label{t2}
\end{table}
\end{document}